\documentclass[12pt,tightenlines,eqsecnum,floats,aps,amsmath,amssymb,nofootinbib,prd,superscriptaddress]{revtex4}

\usepackage{setspace}
\usepackage{amsmath,amssymb,amsfonts,amsthm,mathrsfs}
\usepackage{graphicx}
\usepackage{enumerate} 
\usepackage{comment}
\usepackage{slashed}
\usepackage[pdftex]{hyperref}
\usepackage[normalem]{ulem}
\def\be{\begin{equation}}
\def\ee{\end{equation}}
\def\ba{\begin{eqnarray}}
\def\ea{\end{eqnarray}}
\def\bi{\begin{itemize}}
\def\ei{\end{itemize}}

\def\bra{\langle}
\def\ket{\rangle}
\def\xh{\hat{x}}
\def\qh{\hat{q}}
\def\w{\omega}
\def\G{\mathcal{G}}
\def\lam{\lambda}
\def\A{\mathcal{A}}

\def\I{\mathcal{I}}

\def\H{\mathcal{H}}
\def\t{\tau}

\def\e{\epsilon}

\def\Hp{\mathcal{H}^{+}}
\def\Ho{\mathcal{H}^{o}}

\def\Qs{Q^{\rm soft}}
\def\Qh{Q^{\rm hard}}
\def\Qms{Q^{- \rm soft}}
\def\Qps{Q^{+ \rm soft}}
\def\Qmh{Q^{-\rm hard}}
\def\Qph{Q^{+ \rm hard}}

\def\O{\Omega}

\def\Lint{\mathcal{L}_{\rm int}}

\def\phone{\overset{(1)}{\varphi}}
\def\phn{\overset{(n)}{\varphi}}
\def\lamone{\overset{(1)}{\Lambda}}
\def\asinh{{\rm arcsinh}\,}
\def\Psib{\bar{\Psi}}

\def\dpp{\widetilde{d p}\,}
\def\phis{\varphi_{\mathcal{I}}}
\def\phisp{\varphi_{\mathcal{I}^+}}
\def\phism{\varphi_{\mathcal{I}^-}}
\def\phihp{\varphi_{\mathcal{H}^+}}
\def\phiho{\varphi_{\mathcal{H}^o}}
\def\lamhp{\Lambda_{\mathcal{H}^+}}
\def\lamho{\Lambda_{\mathcal{H}^0}}
\def\lamo{\overset{(-1)}{\lambda}}
\def\lamln{\overset{ (-1,\ln)}{\lambda}}

\def\Lamo{\Lambda^{(1)}}
\def\Lamt{\Lambda^{(2)}}
\def\Lamz{\Lambda^{(0)}}

\begin{document}

\title{Can scalars have asymptotic symmetries?}
%\title{Asymptotic conserved charges from soft scalars}

\author{Miguel Campiglia}
\author{Leonardo Coito}
\affiliation{Instituto de F\'isica, Facultad de Ciencias,  Montevideo 11400, Uruguay}
\author{Sebastian Mizera}
\affiliation{Perimeter Institute for Theoretical Physics, Waterloo, ON N2L 2Y5, Canada}
\affiliation{Department of Physics \& Astronomy, University of Waterloo, Waterloo, ON N2L 3G1, Canada}

\begin{abstract}
Recently it has been understood that certain soft factorization theorems for scattering amplitudes can be written as Ward identities of new asymptotic symmetries. This relationship has been established for soft particles with spins $s > 0$, most notably for soft gravitons and photons. Here we study the remaining case of soft scalars. We show that a class of Yukawa-type theories, where a massless scalar couples to massive particles, have an infinite number of conserved charges. This raises the question as to whether one can associate asymptotic symmetries to scalars.
 %We study the charges in terms of asymptotic fields and explore possible symmetry interpretations. 
   \end{abstract}
%\pacs{}
\maketitle

\tableofcontents

\section{Introduction}

Recently there has been a renewed interest in soft theorems \cite{dass,Weinberg:1965nx,Cachazo:2014fwa,Bianchi:2014gla,Low:1958sn,Burnett:1967km,GellMann:1954kc,Gross:1968in,Jackiw:1968zza,White:2011yy,casali,Mafra:2011nw,Schwab:2014fia,Sen:2017xjn,Cachazo:2015ksa,Cachazo:2016njl,White:2014qia,Klose:2015xoa,Saha:2017yqi,Huang:2015sla,Luna:2016idw,Elvang:2016qvq,Ademollo:1975pf,Schwab:2014sla,DiVecchia:2015oba,Bianchi:2015yta,DiVecchia:2016szw,DiVecchia:2016amo,divecchia,DiVecchia:2015bfa,Bianchi:2016viy,Liu:2014vva,Guerrieri:2017ujb,DiVecchia:2017gfi,DiVecchia:2017uqn} for scattering amplitudes and their relation to asymptotic symmetries \cite{Bondi:1962px,Sachs:1962wk,AS,aajmp,aareview,barnich1,barnich2,bst,ablm,avery}, see \cite{Strominger:2017zoo} for a review. First studies that sparked this activity where in the context of gauge and gravity theories \cite{stromgr,stromym,stromqed,stromst}. It was argued that factorization of an amplitude when one of the momenta becomes soft can be thought of as a consequence of Ward identities for a new set of symmetries acting on asymptotic fields. In the case of gauge theory, this symmetry is generated by large gauge transformations that do not decay at null infinity \cite{stromym,stromqed}. In the case of gravity, they are the Bondi--van der
Burg--Metzner--Sachs supertranslations \cite{Bondi:1962px,Sachs:1962wk}.

Soon after, the ideas were extended in multiple directions, including  higher dimensions \cite{hdqed,hdst}, subleading soft theorems \cite{virasoro,stromlow,clsubgrav,subqed,subsub,conde1,conde2},  massive particles \cite{strommass,clmass,clmass2}, and  other theories \cite{onehalf,threehalfs1,threehalfs2,higherspin1}. For gauge and gravity theories, the soft particle in the soft theorems are of spin-1 (photon) and spin-2 (graviton). In \cite{onehalf} and  \cite{threehalfs1,threehalfs2} the analysis was generalized to the case where the soft particles have spin-1/2 (photino) and spin-3/2 (gravitino) respectively. An analysis for higher spins was given in \cite{higherspin1}. Here, we would like to complete this list by studying the case of a spin-0 (scalar) soft particle.

There are two main motivations for this work. Firstly, there is a large number of theories for which soft theorems have been studied \cite{dass,Weinberg:1965nx,Cachazo:2014fwa,Bianchi:2014gla,Low:1958sn,Burnett:1967km,GellMann:1954kc,Gross:1968in,Jackiw:1968zza,White:2011yy,casali,Mafra:2011nw,Schwab:2014fia,Sen:2017xjn,Cachazo:2015ksa,Cachazo:2016njl,White:2014qia,Klose:2015xoa,Saha:2017yqi,Huang:2015sla,Luna:2016idw,Elvang:2016qvq,Ademollo:1975pf,Schwab:2014sla,DiVecchia:2015oba,Bianchi:2015yta,DiVecchia:2016szw,DiVecchia:2016amo,divecchia,DiVecchia:2015bfa,Bianchi:2016viy,Liu:2014vva,Guerrieri:2017ujb,DiVecchia:2017gfi,DiVecchia:2017uqn}. It is however not clear to what extent a relation to new asymptotic conserved charges can be identified in all these cases. It is then natural to try to map the space of theories for which this identification can be made. We believe this will bring us closer to understanding the nature of asymptotic symmetries in general.  Our second motivation comes from a Minkowski holography approach \cite{deboer,solo} that  has received  fresh attention \cite{sundrum,stromholo} after the soft theorem/asymptotic symmetry developments. The potential use of this approach for scattering amplitudes is being explored in the  simpler setting of  scalar fields \cite{stromholo,cardona}.  We hope that an asymptotic symmetry description of scalar soft theorems could be of help  in this program.

In this paper we study a number of field theories where  a massless scalar field $\varphi$ couples to a massive field $\psi$ through a Yukawa-type interaction,
\be
\Lint \sim \varphi \psi^2  . \label{Lint}
\ee
This provides the simplest example where a soft theorem can be associated to an asymptotic charge. Cubic $\varphi^3$ interactions in four dimensions are in conflict with the asymptotic expansion at null infinity \cite{highdim} and are therefore excluded in our analysis. See also \cite{hamadachiral} for a different --and  phenomenologically more  relevant-- model where soft scalars lead to asymptotic charges.

 For concreteness the field $\psi$ will be taken to be  either a scalar or a Dirac field, but other  fields can be treated similarly. In the theory (\ref{Lint}), the tree-level amplitude of $n$ hard particles of either type, and one soft $\varphi$ scalar factorizes as 
\be
\A_{n+1}(p_1,\ldots,p_n; q) \;\propto\;  \sum_{i \in \psi}^{}\, \frac{1}{  p_i \cdot q}\, \A_{n}(p_1,\ldots,p_n) + \ldots , \label{sthmintro}
\ee
with a proportionality factor that depends on the spin of $\psi$. Hard particles have momenta $p_i$, and the soft one has momentum $q$. The dots denote subleading terms in $q$.

Following the by now standard procedure \cite{stromym}, one can  recast (\ref{sthmintro}) in the form of a Ward identity,
\be
Q^{-}_{\qh} S = S Q^+_{\qh}, \label{wardid}
\ee
for appropriately defined charges $Q^{\pm}_{\qh}$ parametrized by the soft momentum direction $\qh$.  The situation is completely analogous to Ward identities associated to similar soft theorems arising in other theories.  There is however a qualitative difference in theories where  the spin of the soft particle is greater or equal than one: In those cases the charges are associated to large local symmetries. Specifically, the spin-1 soft theorem is associated to  gauge symmetries, the spin-2 theorem to diffeomorphisms, and the spin-3/2 theorem to  local supersymmetry.  The spin zero and  one-half cases  fall in a different category and do not appear to be associated to any underlying local symmetry. 

Even in the absence of local symmetry, the  soft theorem (\ref{sthmintro}) still predicts the existence of conserved charges. Why are they conserved? What is the underlying symmetry?  In this paper we take the first steps towards answering these questions. 

Our analysis will be restricted to tree-level amplitudes. The inclusion of loops would require us  to regard  (\ref{Lint}) as a part of a larger theory where the field $\varphi$ remains massless at quantum level. One possibility could be to realize  $\varphi$ as a `dilaton' (the Nambu-Goldstone boson of spontaneously broken scale symmetry \cite{Callan:1970yg,Boels:2015pta}), which is known  to satisfy the soft theorem  (\ref{sthmintro})  \cite{divecchia}.

The organization of the paper is as follows. In section \ref{wardsec} we describe the soft scalar theorem of interest and rewrite it as in Eq. (\ref{wardid}). %In the next sections we study the charges in (\ref{wardid}) from a classical field theory perspective:  
In section \ref{sec3} we express the charges in terms of the asymptotic fields and establish their conservation as a consequence of the field equations (assuming regular fall-off conditions). In section \ref{secint} we take  first steps towards  unraveling the symmetries underlaying  the charges: We compute their action on asymptotic fields and realize the charges in terms of a spacetime current. We also point out several conceptual difficulties which are similar to ones encountered in `magnetic' charges associated to spin 1 and 2 soft theorems. We conclude in section \ref{sec5} with a summary of  results and  open questions. Some calculations and side comments are left to appendices. \\

%In section  \ref{chargesec} we express the charges in terms of the asymptotic fields and use it  in section  \ref{chargeconssec} to establish charge conservation.  Smeared version of the charges are considered in section \ref{secsm}.   In section \ref{secint} we take the first steps towards understanding the symmetry of underlaying the charges. We compute their action on asymptotic fields, construct a spacetime current, and discuss  similarities and differences with  soft photons and gravitons. 

\noindent Conventions: We use mostly plus signature spacetime metric. The normalization of Fock operators is such that $[a(\vec{p}),a^\dagger(\vec{p}') ] =    (2 \pi)^3 (2 E_p) \delta^3(\vec{p}-\vec{p}')$. We use the following convention for various tensor indices: $a,b,\ldots$ for spacetime abstract indices; $\mu,\nu,\ldots$ for cartesian Minkowski indices; $\alpha,\beta,\ldots$  for  the (space or time-like) 3-hyperboloid and $A,B,\ldots$ for the 2-sphere.

\section{Spin zero soft theorem as a Ward identity} \label{wardsec}
Consider first the case where the  massive  field $\psi$ is a scalar. The coupling with the massless  $\varphi$  is given by the interaction Lagrangian
\be
\Lint = \frac{g}{2} \varphi \psi^2  . \label{Lintpsi}
\ee
A `soft theorem' for this theory  may be established along the same lines as the photon or graviton case. One concludes that the tree-level amplitude of $n$ particles of either type and one soft $\varphi$-particle factorizes as (see appendix \ref{softapp} for the derivation):
\be
\A_{n+1}(p_1,\ldots,p_n; \w q) \stackrel{\w \to 0}{=}  \frac{g}{2 \w} \sum_{i \in \psi} \frac{1}{  p_i \cdot q} \A_{n}(p_1,\ldots,p_n) + O(\w^0) .\label{sthm}
\ee
%(with conventions that momenta are outgoing). 
Following Strominger et.al., it is possible to re-express (\ref{sthm}) in a Ward identity form (\ref{wardid}). If $a(\vec{p})$ and $b(\vec{p})$ are the Fock operators associated to $\varphi$ and $\psi$ respectively, the  appropriate charge is:
\be
Q_{\qh}:= \lim_{\w \to 0}\frac{\w}{2}(a(\w \qh) +a^\dagger(\w \qh)) -\frac{g}{2} \int \dpp \frac{b^\dagger(\vec{p})  b(\vec{p})}{p \cdot q}
\label{charge0}
\ee
where  $\dpp \equiv \frac{d^3 \vec{p}}{ (2 \pi)^3 2 E_p}$ and 
\be
q^\mu=(1,\qh) \label{qmu}
\ee
is the future-pointing null vector associated to the direction $\qh$. To simplify notation we are omitting $\pm$ superscripts.  The charges  $Q^\pm_{\qh}$, acting on the  `out' ($+$) and `in' ($-$)  Fock spaces, have both the form (\ref{charge0}) with the corresponding in/out Fock operators. 

In order to establish the equivalence between the $Q_{\qh}$ Ward identity and the soft theorem one also needs, as in the gauge and gravity cases,  a relation between incoming and outgoing soft particles. In the present case the  relevant relation is
\be
\displaystyle{\lim_{\w \to 0} \w \bra {\rm out} |  S a^\dagger(\w \qh)  | {\rm in} \ket  = - \lim_{\w \to 0} \w \bra {\rm out} | a(\w \qh)  S | {\rm in} \ket }.
\ee
With these ingredients it is straightforward to show the desired equivalence,
\begin{multline}
\bra {\rm out} | [Q_{\qh}, S] | {\rm in} \ket = 0 \iff \lim_{\w \to 0} \w \A_{n+1}(p_1,\ldots,p_n; \w q) =\frac{g}{2} \sum_{i\in \psi} \frac{1}{p_i \cdot q} \A_{n}(p_1,\ldots,p_n) \label{equivalence}.
\end{multline}
% As in the gauge and gravity cases, to show (\ref{equivalence}) one needs, besides the charges  (\ref{charge0}), additional information relating the amplitudes for 
 Following standard nomenclature, we write the charge (\ref{charge0}) as
 \be
 Q_{\qh}= \Qs_{\qh} + \Qh_{\qh} \label{Qhs}
 \ee
where $\Qs_{\qh}$ and $\Qh_{\qh}$ are respectively the terms linear and quadratic in the Fock operators. 
 
Whereas the interest in this paper is in four dimensions, we note that the above expressions are valid in arbitrary spacetime dimensions.

Similar analysis can be repeated in Yukawa theory in four dimensions, where the massive field is of Dirac type.  If we take the interaction Lagrangian as
\be
\Lint= g \varphi \Psib \Psi, \label{Lyuk}
\ee
the analogous soft theorem takes the form (see appendix \ref{softapp})
\be
\A_{n+1}(p_1,\ldots,p_n; \w q) \stackrel{\w \to 0}{=}  \frac{g m}{\w} \sum_{{i \in \{\Psi, \bar{\Psi}\}}} \frac{1}{  p_i \cdot q} \A_{n}(p_1,\ldots,p_n) + O(\w^0) \label{sthmdirac}
\ee
where the sum now is over particles and anti-particles of the $\Psi$ field.  The charge whose Ward identity reproduces the soft theorem is of the form (\ref{Qhs}) with $\Qs_{\qh}$ as before and
\be
\Qh_{\qh}= -g m \sum_s \int \dpp \frac{1}{p \cdot q}\left(b^\dagger_s(\vec{p})  b_s(\vec{p}) + d^\dagger_s(\vec{p})  d_s(\vec{p}) \right)
\label{chargedirac}
\ee
where $b_s$ ($d_s$) are the Fock operators for particles (anti-particles) with spin $s= \pm 1/2$.

\section{Conserved charges} \label{sec3}
\subsection{Charges in terms of asymptotic fields} \label{chargesec}
In this section we write the charges $Q^\pm_{\qh}$ in terms of asymptotic future/past fields. In the following discussion and for the reminder of the paper we restrict attention to four spacetime dimensions.  

The first step is to express the Fock operators in terms of asymptotic fields. For the massless field $\varphi$, this is captured in the null-infinity limit:
\be
r \to \infty \quad \text{with} \quad u=t-r = \text{constant} , \; \xh  = \text{constant}.
\ee
Assuming that in this limit $\varphi(x)$ is given by the free-field expression
\be
\varphi(x) \approx  \int \dpp a(\vec{p}) e^{i p \cdot x} +c.c. 
\ee
 a standard saddle point argument (see e.g. \cite{frolov})  tells
\be
\varphi(x) = \phisp(u,\xh)/r + \ldots \label{phinull}
\ee
with
\be
\phisp(u,\xh) = \frac{1}{4 \pi i}\int_0^{\infty} \frac{dE}{2 \pi} a(\vec{p}=E \xh) e^{-i E u} +c.c.  \label{phia}
\ee
The field $\phis(u,\xh)$ is to be regarded as a field living on future null infinity $\I^+$. From (\ref{phia}) it follows that the `soft' part of the charge $Q^+_{\qh}$ can be written as
\ba
\Qps_{\qh} & = & -4 \pi \int_{-\infty}^{\infty} du \partial_u \phisp(u,\qh) \label{Qsq} \\ 
& = & 4 \pi(\phisp(u=-\infty,\qh) - \phisp(u=+\infty,\qh) ) \label{Qsq2}  . 
\ea
For  massive fields one needs to consider the time-infinity limit 
\be
t \to \infty \quad {\rm with} \quad r/t= {\rm constant}, \; \xh  = \text{constant}.
\ee
It is convenient to write this limit in terms of hyperbolic coordinates 
\be
\t = \sqrt{t^2 - r^2}  ,\quad \rho  =  \frac{r}{\sqrt{t^2 - r^2}} 
\ee
as 
\be
\t \to \infty \quad {\rm with} \quad y^\alpha := (\rho,\xh)= {\rm constant}.
\ee
Assuming again an asymptotic free field expression
\ba
\psi(x) & \approx & \int \dpp  b(\vec{p}) e^{i p \cdot x}   +c.c. \\
\Psi(x) & \approx & \sum_s \int \dpp ( b_s(\vec{p}) u_s(\vec{p}) e^{i p \cdot x} + d^\dagger_s(\vec{p}) v_s(\vec{p}) e^{-i p \cdot x} )
\ea
 a standard saddle point argument tells  (see e.g. \cite{clmass}):
\ba
\psi(x) &= & \frac{\sqrt{m}}{2 (2 \pi \tau)^{3/2}}   b(\vec{p}=m \rho \xh) e^{ -i \t m } + c.c. + \ldots \label{psifock} \\
\Psi(x) & = & \frac{\sqrt{m}}{2 (2 \pi \tau)^{3/2}} \sum_s(   b_s(\vec{p}) u_s(\vec{p}) e^{ -i \t m } +d^\dagger_s(\vec{p}) v_s(\vec{p}) e^{i \t m} )|_{\vec{p}=m \rho \xh} + \ldots. \label{psitifock}
\ea
where we omitted an unimportant overall phase. Similar to the gauge and gravity cases, the `hard' part of the charge will be given by the `source' for the  field $\varphi$. The field equations for $\varphi$ are
\be
\square \varphi =- \frac{g}{2} \psi^2 
\ee
in the scalar $\psi$ theory  and
\be
\square \varphi = - g \Psib \Psi 
\ee
in the Yukawa theory. The  $t \to \infty$ behavior of the massive fields implies that in both cases the  leading term of the source falls as $1/\t^{3}$,\footnote{In the scalar $\psi$ case,  the dots in (\ref{sqphij})  include terms of the form $e^{2 i m \t}/\t^3$, $e^{-2 i m \t}/\t^3$. In the Yukawa case such terms are absents and the dots  start at  $O(\t^{-4})$.} 
\be
\square \varphi = \frac{j(y)}{\t^3}+ \ldots .\label{sqphij}
\ee
In the scalar $\psi$ theory the leading source is given by
\be
j(y) =- g \frac{m}{4 (2 \pi)^{3}}   b^\dagger(\vec{p}) b(\vec{p}) |_{\vec{p}=m \rho \xh} \label{jbb} , 
\ee
whereas in  the Yukawa theory
\be
j(y) = - 2 m g \frac{m}{4 (2 \pi)^{3}} \sum_s \left(b^\dagger_s(\vec{p})  b_s(\vec{p}) + d^\dagger_s(\vec{p})  d_s(\vec{p}) \right)|_{\vec{p}=m \rho \xh}. \label{jyuk}
\ee
$j(y)$ is to be regarded as a field living on the future time infinity hyperboloid $\H^+$  (see appendix \ref{Hpapp}). We now show that  the charge $\Qh_{\qh}$ takes a universal form when written in terms of  $j(y)$.

 Let 
\be
Y^\mu := (\sqrt{1+\rho^2}, \rho \xh) \label{Ymu}
\ee
be the unit time-like vector defined by $y^\alpha$. Under the change of variables $\vec{p} \to y^\alpha:$  $\vec{p}=m \rho \xh$  we have  
\be
p^\mu= m Y^\mu , \quad   \frac{d^3 \vec{p}}{E_p}  =m^2 d^3 V
\ee
with $d^3 V= \sqrt{h} d \rho d^2 \xh$ the volume element on $\Hp$.  Using  this and (\ref{jbb}), (\ref{jyuk}) one finds  that the hard charge in either theory is given by
\be
\Qph_{\qh} = \int d^3 V \frac{j(y)}{Y \cdot q} \label{Qhq}.
\ee
We emphasize that (\ref{Qhq}) is valid for both the scalar $\psi$ or the Yukawa theory. In fact, one of the conclusions one can draw from this paper is that (\ref{Qhq}) is the form of the hard charge regardless the spin of the massive particles.  As  another example in appendix \ref{softapp} we discuss the soft theorem for spin 1 massive particles.  
Most of the remainder of the paper  follows from equations (\ref{Qsq}),  (\ref{sqphij}) and (\ref{Qhq})  and is  thus insensitive to the nature of the massive particles.

Similar analysis applies to the asymptotic past. In terms of advanced time $v=t+r$, the asymptotic form of $\varphi(x)$ 
near  past null infinity is
\be
\varphi(x) = \phism(v,\xh)/r + \ldots \label{phinullm}
\ee
with
\be
\phism(v,\xh) = - \frac{1}{4 \pi i}\int_0^{\infty} \frac{dE}{2 \pi} a(\vec{p}=-E \xh) e^{-i E v} +c.c.  \label{phiam}
\ee
The soft part of $Q^-_{\qh}$ is then:
\be
\Qms_{\qh}  =   4 \pi(\phism(v=+\infty,-\qh) - \phism(v=-\infty,-\qh) ) \label{Qsq2m}  . 
\ee
One can similarly obtain $\Qmh_{\qh}$ in terms of the asymptotic fields at past time infinity. 

\subsection{Charge conservation} \label{chargeconssec}
In the previous section we found  expressions for the charges in terms of the asymptotic fields. The Ward identities discussed  in section \ref{wardsec} tell us  these charges are conserved. The aim of this section is to understand this conservation from the perspective of the classical field theory.

The strategy is as follows.  First, by studying the late-time field equations one can show that
\ba
Q^+_{\qh} & = & 4 \pi  \phisp(u=-\infty,\qh) , \label{Qscripm} \\
Q^-_{\qh} & = & 4 \pi  \phism(v=+\infty,-\qh).  \label{Qscrimp}
\ea
Next, one studies the asymptotic field equations at spatial infinity to show that 
\be
 \phisp(u=-\infty,\qh) = \phism(v=\infty,-\qh) \label{matchphi}
\ee
from which the classical conservation $Q^+_{\qh} = Q^-_{\qh} $ follows. Equation (\ref{matchphi}) is the spin zero version of Strominger's `matching' condition. The  outlined strategy was used in \cite{eyhe} to treat the analogue problem in electrodynamics. The way we link future and past null infinity through spatial infinity is inspired from  \cite{hansen,held,tn}. 

\subsubsection{Field equations at time-infinity} \label{sstiminf}
Let us establish (\ref{Qscripm}). 
From  Eqns. (\ref{Qhs}) and  (\ref{Qsq2})  we see that (\ref{Qscripm}) is equivalent to the condition:
\be
\Qph_{\qh} = 4\pi \phisp(u=+\infty,\qh)  . \label{phiphard}
\ee
To show (\ref{phiphard}) we  study the field equations in the asymptotic time-infinity limit. The field equation  (\ref{sqphij})
 implies  that in this limit
\be
\varphi(x) = \phihp(y)/\t + \ldots \label{phiti}
\ee
with
\be
(D^2+1) \phihp= j. \label{poisson}
\ee
Here $D^2$ is the Laplacian operator on the time-infinity hyperboloid $\H^+$ (see appendix \ref{Hpapp}). Thus,  (\ref{poisson}) is a Poisson-type equation on $\H^+$ that determines $\phihp$ for a given source $j$. The solution can be given in terms of appropriate Green's function,
\be
\phihp(y) = \int d^3 V' \G(y;y') j(y'). \label{solpoiss}
\ee
As shown in appendix \ref{Hpapp}, the properties of this Green's function imply:
\ba
\lim_{\rho \to \infty} \rho \, \phihp(\rho,\qh) & = &  \frac{1}{4 \pi}  \int d^3 V \frac{j(y)}{Y \cdot q} \\
&=& \frac{1}{4 \pi} \Qh_{\qh}.
\ea
Finally,    consistency  between the expansions at null (\ref{phinull}) and time (\ref{phiti}) infinities imply the following continuity condition:\footnote{See \cite{eyhe} for a discussion in the context of electrodynamics.} 
\be
\lim_{\rho \to \infty} \rho \, \phihp(\rho,\xh) = \phisp(u=+\infty,\xh),\, \label{nulltilim}
\ee
from which  (\ref{phiphard}) (and hence (\ref{Qscripm}))  follows. 

Similar treatment applies to asymptotic past. The analogue of relation (\ref{phiphard}) in this case is:
\be
\Qmh_{\qh} = 4\pi \phism(v=-\infty,-\qh)  ,
\ee
which together with (\ref{Qsq2m})  leads to Eq. (\ref{Qscrimp}).
\subsubsection{Field equations at spatial infinity}
In order to relate the future and past charges we now look at the field $\varphi$ near spatial infinity. Hyperbolic coordinates adapted to this end are 
\be
\rho:= \sqrt{r^2 -t^2}, \quad \tau := \frac{t}{\sqrt{r^2-t^2}}.
\ee
The behavior of regular massless scalar at $\rho \to \infty$ is \cite{beig},
\be
\varphi(x) = \phiho(y)/\rho + \ldots
\ee
whereas the massive field $\psi$ falls-off as $e^{- m \rho}$. Thus asymptotically $\varphi$ satisfies the free wave equation, which in turn implies $\phiho(y)$ satisfies
\be
(D^2-1) \phiho=0 \label{weqphone}
\ee
where $D^2$ is the wave operator in the space-infinity hyperbolid $\Ho$ (see appendix \ref{Hoapp}). Equation (\ref{weqphone}) can be seen to imply either $O(1/\t)$ or $O(\ln \t / \t)$ fall-off at $\t \to \infty$. The latter however is not consistent with the assumed behavior at infinity. In fact, a   continuity argument as in Eq. (\ref{nulltilim}) tells
\ba
\lim_{\t \to \infty} \t \phiho(\t,\xh) &= & \phisp(u=-\infty,\xh) ,  \label{limspinul} \\
\lim_{\t \to - \infty} -\t \phiho(\t,\xh) &= & \phism(v=+\infty,\xh).
\ea
For future use, let   $\phi_{\pm}(\xh)$ denote the above asymptotic values, that is,
\be
\phiho(\t,\xh) \stackrel{\t \to \pm \infty}{\to} = \frac{1}{|\t|} \phi_{\pm}(\xh) .\label{limphone}
\ee
In appendix  (\ref{Hpapp}) it is shown these asymptotic values satisfy
\be
\phi_-(\xh)= \phi_+(-\xh) ,  \label{antipphi}
\ee
as a consequence of the wave equation (\ref{weqphone}). This in turn implies Eq. (\ref{matchphi}) and hence the  charge conservation  $Q^+_{\qh}=Q^-_{\qh}$.

\subsection{Smeared charges} \label{secsm}
In analogy to the gauge and gravity cases we now consider smeared version of the charges,
\be
Q^+[\lambda]   :=  \frac{1}{4 \pi} \int_{S^2} d^2 \qh \lambda(\qh) Q^+_{\qh}, \label{Qlam} 
\ee
where $\lambda(\qh)$ is an arbitrary function on the sphere.  
Using the splitting $Q^+_{\qh}= \Qps_{\qh}+\Qph_{\qh}$ and Eqns. (\ref{Qsq}), (\ref{Qhq}) we write $Q^+[\lambda]$ as a sum of soft and hard pieces with
\ba
\Qps[\lambda] & =& - \int_{\I^+} du d^2\xh \lambda(\xh) \partial_u \phisp(u,\xh) , \label{Qpsl} \\
\Qph[\lambda]  & =&  - \int_{\Hp} d^3V \lamhp(y) j(y) , \label{Qphl}
\ea
where we defined:
\be
\lamhp(y) :=- \frac{1}{4 \pi} \int d^2\qh \frac{\lambda(\qh)}{q \cdot Y}. \label{lamone}
\ee

As described in appendix \ref{Hpapp},  definition (\ref{lamone}) implies $\lamhp$ is the function on $\Hp$  that satisfies the Laplace-type equation
\be
(D^2+1) \lamhp =0 \label{weqlam}
\ee
and has the $\rho \to \infty$ behavior $\lamhp(\rho,\xh) \sim \frac{\ln \rho}{\rho} \lambda(\xh)$. Equation (\ref{weqlam})  suggests   $\lamhp$ should be interpreted as  the time-infinity asymptotic value of a spacetime field
\be
\Lambda(x) = \lamhp(y) / \t + \ldots \label{LamHp}
\ee
that satisfies (asymptotically) the free wave equation  $\square \Lambda=0$.\footnote{Another possibility leading to (\ref{weqlam})  would be $\square \Lambda=0$ with $\Lambda(x) \sim \ln \t /\t \lamhp(y)$. However, this option  does not seem to  yield any sensible spacetime picture for the smeared charges.}   A simple expression for a spacetime field satisfying these conditions is obtained by diving Eq.  (\ref{lamone}) by $\t$:
\be
\Lambda(x) := - \frac{1}{4 \pi} \int d^2\qh \frac{\lambda(\qh)}{q \cdot x} \label{lambulk}.
\ee
Indeed one can verify such expressions satisfies $\square \Lambda=0$  and Eq. (\ref{LamHp}) with no subleading terms.  We leave for next section further discussion on this spacetime perspective. %This  spacetime perspective will be explored in the next section. 

We now go to the spatial infinity description of (\ref{Qlam}).  The spacetime field $\Lambda(x)$ introduced above has a spatial infinity expansion,
\be
\Lambda(x) = \lamho(y) / \rho + \ldots \label{LamHo}
\ee
with $\lamho$ satisfying the wave equation
\be
(D^2-1)\lamho =0.
\ee
  From this  perspective,  $\lambda$  arises  as the  asymptotic value of  $\lamho$ on $\Ho$,
\be
 \lamho \stackrel{\t \to \infty}{\sim} \frac{\ln \t}{\t} \lambda(\xh). \label{limlamone}
\ee
The structure is  analogous  to the gauge and gravity cases. There, the symmetry parameter on $\Ho$ ($\lamho(y)$ in the present case)  satisfies the same differential equation as the field associated to the unsmeared charge ($\phiho(y)$ in the present case). The charge is then written as a Klein-Gordon symplectic product between the two fields and is thus conserved. In the present case, this amounts to define the following conserved charge:
\be
Q_\t = \int_{\t={\rm const.}} dS_\alpha \sqrt{h}h^{\alpha \beta}(\partial_\beta \phiho \lamho-\partial_\beta \lamho \phiho) .\label{Qtau}
\ee
The integral (\ref{Qtau}) can be evaluated in the $\t \to \infty$ limit by using the fall-offs (\ref{limphone}) and (\ref{limlamone}). Doing so  one finds:
\be
Q_{\t = \infty}  = \int_{S^2} d^2 V \lambda(\xh) \phi_+(\xh),
\ee
which precisely coincides with the charge (\ref{Qlam}).   By taking the $\t \to -\infty$ limit one obtains charge conservation between the asymptotic past and future smeared charges.

\section{Symmetry interpretation of charges?} \label{secint}
In the previous sections we expressed the scalar soft theorem (\ref{sthmintro}) as the conservation of  certain charges defined at the asymptotic past and future, $Q^-[\lambda]=Q^+[\lambda]$. These charges are spin zero analogues of the charges associated to spin 1 and 2 soft theorems, schematically:
\ba
Q_{s=0} & \sim & \int_\I  \lambda \partial_u \phi + \ldots \label{s0Q} \\
Q_{s=1} & \sim &  \int_\I  D^A \lambda \partial_u A_A + \ldots  \label{s1Q}\\
Q_{s=2} & \sim &  \int_\I  D^A D^B \lambda \partial_u C_{AB} + \ldots \label{s2Q},
\ea
where we are only displaying the `soft' part of the charges and $\phi$, $A_A$, $C_{AB}$ represent the scalar, photon and graviton field at null infinity. The analogy between the $s=0$ and $s=1,2$ charges appears to stop when it comes to symmetries.  Expressions (\ref{s1Q}) and (\ref{s2Q}) can be understood as canonical charges associated to `large' $U(1)$ gauge symmetries $\lambda(\xh)$ and `supertranslation' diffeomorphisms $\xi_\lambda= \lambda(\xh) \partial_u$ respectively, but there is no obvious  symmetry interpretation for (\ref{s0Q}). A related point is the fact that for $s=1,2$, $Q_s$ includes `global' charges: total electric charge and total linear momentum respectively, whereas there is no global charge associated to (\ref{s0Q}). 

In the spin 1 and 2 cases there are in fact a second class of charges associated to soft theorems: These are the `magnetic'  versions of (\ref{s1Q}) and (\ref{s2Q}). For instance, in the $s=1$ case it is given by \cite{strommag,subqed}
\be
\widetilde{Q}_{s=1}  \sim   \int_\I  \e^{AB} D_A \lambda \partial_u A_B, \label{s1Qt}
\ee
where $\e^{AB}$ is the area 2-form on the sphere. These magnetic charges have a closer analogy with the $s=0$ charges in that they do not include `global' charges\footnote{Unless one allows for magnetic monopoles \cite{strommag}.} and their symmetry description  is subtle. A simple symmetry interpretation of (\ref{s1Qt}) can be given if one works with `dual' variables $\tilde{A}_A$ describing the potential of the dual field strength $\tilde{F}_{\mu \nu} \sim \e_{\mu \nu \rho \sigma}F^{\rho \sigma}$ \cite{strommag,hamadaem}.\footnote{We thank Beatrice Bonga, Laurent Freidel and Ali Seraj for discussions on this point.\label{thxbs}} In the scalar case, we do not know of an alternative description that would make the symmetry associated to (\ref{s0Q}) transparent. As a first step towards understanding this symmetry, in subsection \ref{ss1} we  compute the action of the charges on the asymptotic data.\\ %We will do so in the next subsection. However, we do not have a clean geometrical interpretation of this action. If the magnetic charge example is to be of any guidance, there could be an alternative description where the transformation becomes more transparent.

Another aspect of the charges associated to soft theorems is that they can be written as spatial integrals of total derivative currents (see e.g. \cite{avery}), 
\be
j^a = \partial_b k^{a b} \label{jk}
\ee
where  $k^{ab}$ is a densitized antisymmetric tensor. For instance, in the electrodynamics case we have $k^{a b} = \sqrt{\eta} \Lambda F^{ab}$ and $k^{a b} = \sqrt{\eta} \Lambda \tilde{F}^{ab}$ for the electric and magnetic charges respectively, where $\Lambda(x)$ is the `large gauge' parameter such that  $\Lambda(x) \to \lambda(\xh)$ at null infinity. In subsection \ref{ss2} we describe a current (\ref{jk}) that reproduces the charge (\ref{s0Q}).

\subsection{Action of charges on asymptotic fields} \label{ss1}
From the perspective of the asymptotic future, the smeared charge $Q^+[\lambda]$ (\ref{Qlam}) is defined on the space of asymptotic free fields. One can thus compute its action on the asymptotic fields by Poisson brackets/commutators.  For the field $\varphi$ the relevant term is the `soft' charge  (\ref{Qpsl}) from which one finds,
\be
\delta_{\lambda} \phisp(u,\xh) = -\lambda(\xh), \label{dellamphisp}
\ee
Thus, asymptotically the field $\varphi$ transforms by a `shift'
\be
\delta_{\lambda} \varphi = -\lambda(\xh)/r + \ldots \label{dellamphi}
\ee
where the dots denote terms subleading in $1/r$. %\footnote{One may wonder whether the shifting field in (\ref{dellamphi})  corresponds to the spacetime field $\Lambda(x)$ introduced in Eq. (\ref{LamHp}). The answer is in the negative. As shown in appendix \ref{lamapp}, $\Lambda(x)$ has a leading $\ln r /r$  behavior  at null infinity. This apparent mismatch  is discussed further in subsection \ref{ss3}.}

For the  field $\psi$, it is  easier to first study the transformation on the asymptotic Fock operator $b(\vec{p})$. The relevant piece is now the `hard' charge (\ref{Qphl}) from which one concludes
\be
\delta_{\lambda} b(\vec{p})= - i \frac{g}{2m} \lamhp(\vec{p}/m) b(\vec{p}). \label{delb}
\ee 
Given the asymptotic form of $\psi$ in terms of $b$ (\ref{psifock}), we  can   infer the asymptotic action on $\psi$. We seek an  expression involving only spacetime quantities.  Let us interpret  $\lamhp$ in (\ref{delb}) as the  $\t^{-1}$ coefficient of the  spacetime field $\Lambda(x)$ (\ref{LamHp}). The appearance of an $i$ in (\ref{delb}) suggests a time-derivative action on $\psi$. In order to compensate for the $\t^{-1}$ term, we are lead to consider the time-derivative vector field:
\be
X^a= \t \partial_\t, \label{XHp}
\ee
which is just the dilatation vector field $x^\mu \partial_\mu$ expressed  in hyperbolic coordinates.  With these ingredients one concludes: 
\be
\delta_{\lambda} \psi = \frac{g}{2m^2} \Lambda X^a \partial_a \psi + \ldots \label{deltalampsi}
\ee
where the dots denote terms subleading in $1/\t$. 

As stated in the beginning of the section, we do not have a geometric understanding for the asymptotic transformation (\ref{dellamphisp}), (\ref{deltalampsi}).  The transformation for the field $\varphi$ suggests a `shift' transformation analogous to the one occurring in electromagnetic or gravity cases. The analogy however fails in describing the `total derivative' nature of the current (\ref{jk}), see appendix \ref{ss3} for details. The appearance of the dilatation vector field in (\ref{deltalampsi}) suggests a relation to scale symmetry but we have not been able to establish any concrete connection.\footnote{This point could perhaps be clarified in an a setting where $\varphi$ is a dilaton field. See the discussion section for further comments.}

Regardless of the interpretation, the above transformation suffers a basic problem we now describe. If we think of the future asymptotic data as given by $\phisp(u,\xh)$ and $b(\vec{p})$, it is in fact not entirely free but must satisfy Eq. (\ref{phiphard}) as described in section \ref{sstiminf}. This condition is  violated by the transformation (\ref{dellamphisp}), (\ref{delb}), since it changes $\phisp(u=\infty,\xh)$ while leaving $\Qph_{\qh}$ unchanged. This problem hints at the need to include boundary terms in the symplectic product. We note that similar problem occurs for the `magnetic' shifts generated by (\ref{s1Qt}), which violates the condition $\e^{AB} \partial_A A_B(u=\infty,\xh)=0$ that electromagnetic fields satisfy in the absence of magnetic monopoles \cite{winicour}.\textsuperscript{\ref{thxbs}}

\subsection{Spacetime current} \label{ss2}

Our aim is to find an antisymmetric  $k^{ab}$ whose current $j^a= \partial_b k^{ab}$   reproduces the smeared charges of section \ref{secsm}.   Note that this requirement only determines $k^{ab}$  asymptotically but otherwise  leaves it arbitrary in the bulk. Below we provide a $k^{ab}$ that  has a particularly simple spacetime form.

As in the gauge and gravity case, we seek for a $k^{ab}$ that depends on the massless field $\varphi(x)$ and on the ``symmetry parameter'' field $\Lambda(x)$ introduced in the previous section.  We will also need a third ingredient, the dilatation vector field
\be
X^a = x^\mu \partial_\mu.
\ee
In terms of $\varphi$, $\Lambda$ and $X^a$ we define $k^{ab}$ by:\footnote{An equivalent form is $k^{ab}= 2\sqrt{\eta}\left( \Lambda \nabla^{[a} \varphi X^{b]} - \varphi \nabla^{[a} \Lambda X^{b]} \right)$ where we used that  $\nabla^{[a}X^{b]}=0$.}
\be
k^{ab}= \sqrt{\eta}\left( (\nabla^a \varphi \Lambda - \nabla^a \Lambda \varphi)X^b - (a \leftrightarrow b) \right). \label{kab}
\ee

We now show this current reproduces the smeared charges. Let us start by discussing the charges form the spatial infinity perspective. Consider the hyperbolic coordinates $(\rho,\t,\xh)$ adapted to spatial infinity. In these coordinates, the dilation vector field  takes the form
\be
X^a = \rho \partial_\rho.
\ee
Take  a spacetime Cauchy slice $\Sigma_\t$ that approaches a $\t=$const. slice as $\rho \to \infty$. The integral of $j^a$ over $\Sigma_\t$ is then given by
\be
\int_{\Sigma_\t} d S_a \partial_b k^{a b} =  \lim_{\rho \to \infty} \int_{\t={\rm const.}}d^2 \xh k^{\tau \rho} .\label{intjsigma}
\ee
Using the $\rho \to \infty$ expansion of the fields one finds,
\be
k^{\tau \rho} = \sqrt{h}h^{\t \t}(\partial_\t \phiho \lamho -\partial_\t \lamho \phiho) + O(\rho^{-1}).
\ee
and one recovers the charge $Q_\t$ given in Eq.  (\ref{Qtau}).

We now discuss the charge from the perspective  of the asymptotic future. Consider a family of Minkowski time $t=$constant slices and evaluate the integral of $j^a$ in a $t \to \infty$ limit:
\be
Q^{\infty} := \lim_{t \to \infty} \int_{t = {\rm const.}} dS_a  j^a. \label{Qfut}
\ee
Since $j^a$ is built out of both massless and massive fields, there appears two contributions to (\ref{Qfut}) one associated to null infinity and the other to  time-like infinity \cite{clmass}. To study the null infinity contribution we go to  retarded $(r,u)$ coordinates. In these coordinates the dilatation vector field reads
\be
X^a= r \partial_r + u\partial_u, \label{Xru}
\ee
$\varphi$ has the expansion (\ref{phinull}) and $\Lambda(x)$ the expansion (see appendix \ref{lamapp})
\be
\Lambda(r,u,\xh) = \frac{\ln r}{2 r} \lam(\xh) - \frac{1}{2 r}  \ln |2 u| \lam(\xh) + O(r^{-1-\epsilon}). \label{lamnull}
\ee
From these expressions one can evaluate  (\ref{Qfut}) in the $t \to \infty, u=$const. limit to obtain the null infinity contribution to the charge:
\ba
Q^{\I^+} & := &\lim_{t\to \infty}  \int du d^2 \xh \partial_u k^{ru} \label{qnullfirst} \\
& = & - \int du d^2 \xh \lam(\xh) \partial_u \phisp(u,\xh), \label{qnullfin}
\ea
which recovers (\ref{Qpsl}). Note that the result is not entirely obvious  since there are potentially  logarithmic divergent terms in (\ref{qnullfirst}) that either cancel out or integrate to zero. See appendix \ref{lamapp} for details.

The time-like infinity contribution to (\ref{Qfut}) is given by evaluating the integral in the limit where $t \to \infty$ with $r/t=$const. As before, this limit is most conveniently described in  hyperbolic coordinates adapted to time infinity. In this coordinates the dilatation vector field is  given by Eq. (\ref{XHp}). 
The time-like infinity contribution is found to be  given by:
\ba
Q^{\Hp} & := &\lim_{\t\to \infty}  \int_{\t = {\rm const.}} \partial_\alpha k^{\t \alpha}\\
& = & - \int_{\Hp} \partial_\alpha \left(\sqrt{h} h^{\alpha \beta} (\partial_\beta \phihp \lamhp - \partial_\beta \lamhp \phihp) \right) \label{qspibdy}\\
& = & - \int_{\Hp} d^3V  \lamhp  j . \label{qspifin}
\ea
The second line follows from the time-infinity fall-offs and the third line from the field equations at time-infinity. Expression (\ref{qspifin}) recovers  (\ref{Qphl}).\footnote{One can also get this result by evaluating (\ref{qspibdy}) in terms of the $\rho \to \infty$ asymptotics of $\phihp$ and $\lamhp$.}  

In this way we see how the current $j^a=\partial_b k^{ab}$  reproduces the expected charge in the asymptotic future. Similar result  applies to the asymptotic past charge.

\section{Discussion} \label{secsummary} \label{sec5}
In this work we studied a family of soft factorization theorems for scalar particles and showed their equivalence to Ward identities of an infinite number of new asymptotic charges. We identified the charges in terms of asymptotic fields and proved their conservation as a consequence of the field equations. As in other soft theorems, we showed the charges can be written in terms of a total derivative current. 

The main open question regards the asymptotic symmetry interpretation of the charges. As a first step in answering this question, we computed the action of the charges on the asymptotic fields. However, we found a conflict between the resulting transformations and a constraint occurring in the asymptotic fields. We hope to resolve this issue in future work.

An interesting set up where to  explore the question of symmetry interpretation could be to realize the massless scalar  as the dilaton field of spontaneously broken conformal symmetry. As shown in \cite{divecchia}, the Ward identities of this broken symmetry imply  a soft theorem of the type studied here.\footnote{An analogous  derivation of the soft photon theorem is  given in \cite{Ferrari:1971at,Mirbabayi:2016xvc}. See \cite{Guerrieri:2017ujb,DiVecchia:2017uqn} for recent  discussions on deriving soft theorems from Ward identities of spontaneously broken  symmetries.}  It would also be interesting to study the dilaton subleading soft theorems  \cite{divecchia} from the perspective of asymptotic symmetries.

%@@@@@@comentar derivacion de photon theorem a partir de spont broken symmetry y referir al apendice por analogo aca@@@@@@@

\begin{acknowledgments}
M.C. would like to thank Alok Laddha for insightful discussions and for suggesting studying the action of  charges on  asymptotic fields. S.M. would like to thank Freddy Cachazo and Andrew Strominger for useful discussions. We thank the organizers and participants of the Perimeter Institute workshop `Infrared Problems in QED and Quantum Gravity' for creating a stimulating environment. We thank Beatrice Bonga, Laurent Freidel, Florian Hopfmueller, Ali Seraj and Ronak Soni for stimulating discussions. We thank an anonymous referee for valuable feedback.  This research was supported in part by Perimeter Institute for Theoretical Physics. Research at Perimeter Institute is supported by the Government of Canada through the Department of Innovation, Science and Economic Development Canada and by the Province of Ontario through the Ministry of Research, Innovation and Science.
\end{acknowledgments}

\appendix

\renewcommand{\thefigure}{\thesection.\arabic{figure}}

\section{Soft scalar theorems} \label{softapp}

\begin{figure}[!h]\label{fig:softtopologies}
\centering
\includegraphics[width=0.2\linewidth]{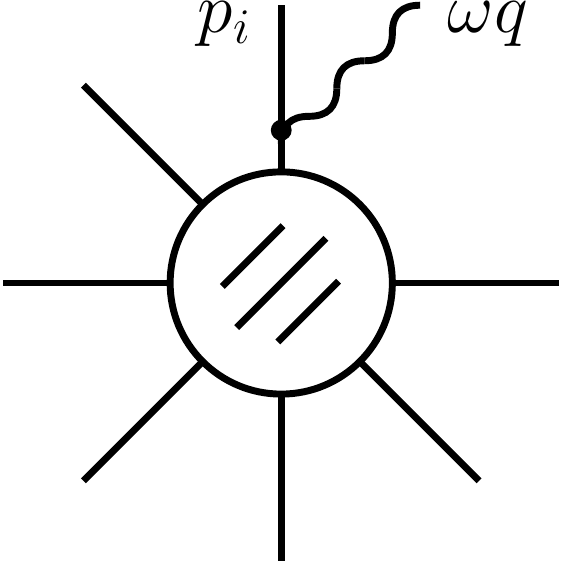}\hspace{5em}
\includegraphics[width=0.2\linewidth]{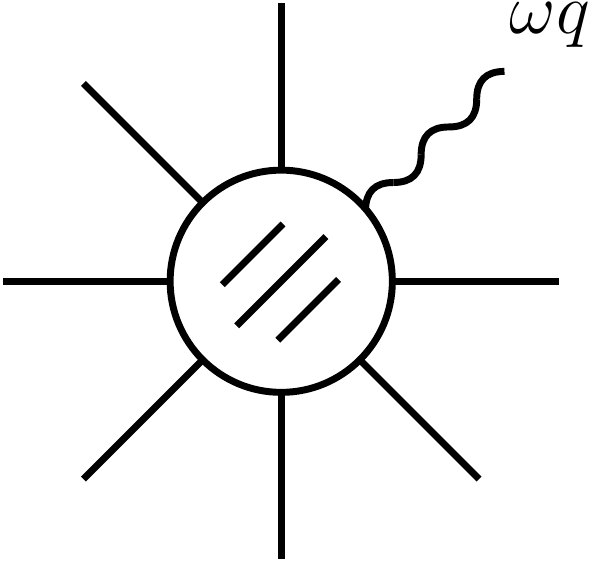}
\caption{Two topologies contributing in the soft limit. Wavy line denotes the soft particle, while straight lines denote the hard ones. All momenta are incoming. Only the left topology contributes to the leading soft behavior.}
\end{figure}

In this appendix we derive the soft theorems discussed throughout this work. We follow the classic derivation due to Weinberg \cite{Weinberg:1965nx,Weinberg:1995mt} using a Feynman diagram argument. For any local theory, an external soft particle can couple to the remaining ones in two different ways, see. Fig. \ref{fig:softtopologies}.

Theories considered in this work have a massless field $\varphi$ coupled via a three-valent vertex to a massive field.\footnote{Note that we do not have to make any assumptions about self-interactions of the massive field, and we allow self-interactions of the massless field $\varphi$ starting at quartic order.} In these cases, the first class of diagrams behaves as $\mathcal{O}(q^{-1})$, yielding the leading contribution to the soft factor. Let us consider in turn theories with different spin of the massive field.

First, we study the spin-zero case given by the interaction Lagrangian:
\be\label{Lint1}
\mathcal{L}_{\mathrm{int}} = \frac{1}{2} g \varphi \psi^2,
\ee
where $\psi$ is a massive real scalar field. Let us study a tree-level scattering amplitude with $n+1$ particles, where one massless scalar with incoming momentum $\omega q$ becomes soft, i.e., $w \to 0$. In this limit, the contribution from coupling to an external massive particle $\psi$ with incoming momentum $p_i$ is
\be
ig\; \frac{-i}{(p_i + \omega q)^2 + m^2}\; \tilde{\mathcal{A}_n} (p_1, \ldots, p_i + \omega q,\ldots, p_n),
\ee
where the intermediate propagator corresponds to a massive particle due to the nature of the interaction term \eqref{Lint1}. Here, $\tilde{\mathcal{A}}_n$ represents the rest of the amplitude. In the soft limit, $\omega \to 0$, we obtain
\be
\frac{g}{2\omega}\frac{1}{p_i \cdot q}\; \tilde{\mathcal{A}_n} (p_1, \ldots, p_i, \ldots, p_n),
\ee
where now $\tilde{\mathcal{A}}_n$ becomes an on-shell amplitude for all particles except for the soft one. Summing over all diagrams contributing to this process gives us the soft factorization:
\be
\mathcal{A}_{n+1} (p_1, \ldots, p_n; \omega q)\; \stackrel{\w \to 0}{=}\; \frac{g}{2 \omega} \sum_{i \in \psi}\frac{1}{p_i \cdot q}\; \mathcal{A}_n (p_1, \ldots, p_n) + \mathcal{O}(1),
\ee
where the sum proceeds over all external massive particles $\psi$, and the subleading terms come from diagrams of different topology.

Having outlined a general strategy, we can now generalize to a spin-half case. We study Dirac fermions in four dimensions coupled to the massless scalar via Yukawa interaction,
\be\label{Lint2}
\mathcal{L}_{\mathrm{int}} = g \varphi \bar{\Psi} \Psi.
\ee
A contribution from an incoming particle $\Psi$ becomes
\be
\tilde{\mathcal{A}_n} (p_1, \ldots, p_i + \omega q,\ldots, p_n)\, i g\, \frac{-i(-\slashed{p}_i-\omega \slashed{q} + m)}{(p_i + \omega q)^2 + m^2}\, u_{s_i}(\vec{p_i}).
\ee
In the soft limit $\omega \to 0$ we find:
\be
\tilde{\mathcal{A}_n} (p_1, \ldots, p_i,\ldots, p_n)\, \frac{g}{2\omega} \frac{-\slashed{p}_i + m}{p_i \cdot q}\, u_{s_i}(\vec{p_i}).
\ee
We now use the spinor identities,
\be
\sum_{s'} u_{s'}(\vec{p_i}) \bar{u}_{s'}(\vec{p_i}) = -\slashed{p}_i + m, \qquad\qquad \bar{u}_{s'}(\vec{p_i}) u_{s_i}(\vec{p_i}) = 2 m \delta_{s' s_i},
\ee
to obtain the soft contribution:
\be
\frac{g m}{\omega} \frac{1}{p_i \cdot q}\, \tilde{\mathcal{A}_n} (p_1, \ldots, p_i,\ldots, p_n)\, u_{s_i}(\vec{p_i}),
\ee
where the spinor $u_{s_i}(\vec{p_i})$ is now decorating the particle which used to give the propagator. Note that it has the same spin and momentum as the original particle $p_i$. Similar expression can be obtained for an incoming anti-particle $\bar{\Psi}$, and outgoing particles using crossing symmetry. After summing over all contributions, we find the soft factorization:
\be
\mathcal{A}_{n+1} (p_1, \ldots, p_n; \omega q)\; \stackrel{\w \to 0}{=}\; \frac{g m}{\omega} \sum_{i \in \{\Psi, \bar{\Psi}\}}\frac{1}{p_i \cdot q}\; \mathcal{A}_n (p_1, \ldots, p_n) + \mathcal{O}(1).
\ee
The leading soft factor is proportional to the mass of the fermion. Since the limit $m \to 0$ is smooth, we conclude that the leading soft factor in a massless Yukawa theory vanishes.

Finally, let us consider the spin-one case. We couple the scalar to massive Z-bosons with the interaction term
\be\label{Lint3}
\mathcal{L}_{\mathrm{int}} = \frac{1}{2} g \varphi Z_\mu Z^{\mu}.
\ee
Contribution from a single diagram for an incoming Z-boson with momentum $p_i$ and polarization $\lambda_i$ is
\be
ig\; \frac{-i(\eta_{\mu\nu} + (p_i + \omega q)_\mu (p_i + \omega q)_\nu / m^2)}{(p_i + \omega q)^2 + m^2}\, \epsilon^{\mu}_{\lambda_i}(p_i)\; \tilde{\mathcal{A}}_n^{\nu} (p_1, \ldots, p_i + \omega q,\ldots, p_n).
\ee
Using the identities,
\be
\sum_{\lambda'=\pm,0} \epsilon_{\lambda'}^{\mu\ast}(p_i) \epsilon_{\lambda'}^{\nu}(p_i) = \eta^{\mu\nu} + p_i^\mu p_i^\nu / m^2, \qquad \epsilon_{\lambda'}^{\ast}(p_i) \cdot \epsilon_{\lambda_i}(p_i) = \delta_{\lambda' \lambda_i},
\ee
and a derivation analogous to the previous cases, we find that in the soft limit,
\be
\mathcal{A}_{n+1} (p_1, \ldots, p_n; \omega q)\; \stackrel{\w \to 0}{=}\; \frac{g}{2\omega} \sum_{i \in Z}\frac{1}{p_i \cdot q}\; \mathcal{A}_n (p_1, \ldots, p_n) + \mathcal{O}(1).
\ee

The discussion in this appendix has been restricted only to tree-level amplitudes. Unlike in the case of soft theorems studied so far \cite{stromym,stromqed,stromst,onehalf,threehalfs1,threehalfs2}, at loop level the massless scalar considered in this work acquires mass. In this case, one cannot take a soft limit, meaning that the asymptotic symmetries identified in this work must be broken on a quantum level. However, it might be possible to introduce extra fields into the theories under considerations, such that an enhanced symmetry would protect masslessness of $\varphi$ at loop level.

\section{Green's functions on $\Hp$, $\Ho$} \label{app}
In these appendix we describe the differential operators and associated Green's functions that arise when expanding the fields at time and spatial infinity.  Most of the considerations are particular cases of the general analysis given in  \cite{deboer}. But there is also some new (to us) material,  for instance sections \ref{pde1Hoapp} and \ref{relHoapp}.
\subsection{(Future) Time-like infinity $\H^+$} \label{Hpapp}
Here we work in hyperbolic coordinates adapted to future time infinity 
\be
\t= \sqrt{t^2-r^2}, \quad \rho = \frac{r}{\sqrt{t^2-r^2}},
\ee
in terms of which the Minkowski line element reads:
\be
ds^2= -d \t^2 + \t^2 d \sigma^2
\ee
where
\be
d \sigma^2 = \frac{d \rho^2}{1+\rho^2}+\rho^2 q_{AB} dx^A dx^B \equiv h_{\alpha \beta} d y^\alpha d y^\beta
\ee
is the metric of the  future unit hyperboloid  $\Hp$.

For a field of the form
\be
\varphi(x) = \t^{-n}\phn(y)
\ee
one has 
\be
\square \varphi(x) = \t^{-n-2}(D^2 \phn+n(2-n) \phn), \label{boxtn}
\ee
where $D^2 \equiv h^{\alpha \beta}D_\alpha D_\beta$ is the Laplacian on $\Hp$. %We will be interested in understanding properties of the differential operator on $\Hp$ defined by (\ref{boxtn}). %We will consider the case of real (in fact integer) $n$ (see \cite{}) for more general discussion. In such case one can verify that
By studying the differential operator on $\Hp$ defined by (\ref{boxtn})  one concludes:
\be
(D^2 +n(2-n) )\phn =0 \; \implies \; \phn  \sim 1/\rho^n \quad {\rm or} \quad  \phn \sim 1/\rho^{2-n}  \label{D2Hp}
\ee
when $\rho \to \infty$. For the present paper we are interested in the $n=1$ case. This happens to be the special  case where the two independent asymptotic solutions  (\ref{D2Hp}) coincide. What occurs then  is that a second independent solution appears with a $\ln \rho$ dependence \cite{deboer}:
\be
(D^2 +1) \phone =0 \; \implies \; \phone \sim 1/\rho \quad {\rm or} \quad  \phone \sim \ln \rho/\rho  . \label{freeHp}
\ee
The PDE  problems appearing in our discussion are:\footnote{The source $j(y)$ decays fast enough at $\rho \to \infty$ so that the asymptotic form of $\phone$ is still dictated by the source-free equation (\ref{freeHp}). The   absence of $\ln \rho/\rho$ term in $\phone$ is due to the regular behaviour of $\varphi$ at null infinity. It is also interesting to note that fields $f$ such that $(D^2+1) f=0$ with $f=O(1/\rho)$ are always singular at $\rho=0$ (this can be established by writing the explicit solution in terms of  spherical harmonics).} 
\ba
(D^2 +1)\phone  &=  j  & {\rm with}  \quad \phone  \sim 1/\rho \label{pde1Hp} \\
(D^2 +1)\lamone & = 0 &{\rm with} \quad \lamone \sim \frac{\ln \rho}{\rho}\lambda(\xh) . \label{pde2Hp}
\ea
We now describe the Green's function associated to  each equation
\subsubsection{Green's function for Eq. (\ref{pde1Hp})}
We want to find the solution to 
\be
(D^2 +1)\G(y;y') =\delta^{(3)}(y,y').
\ee
 Due to the $SO(3,1)$ symmetry it has to be of the form
 \be
 \G(y;y') = g( Y \cdot Y')
 \ee
for some function $g$. We find the general solution for $y \neq y'$ and then get the overall coefficient by requiring the appropriate  $y \to y'$ behavior. Taking $\rho'=0$ so that $Y \cdot Y'=- \sqrt{1+\rho^2}$  one finds the general solution to $(D^2 +1)f=0$ is given by
%(D^2 +1)f=  (P^2-1)f''(P)+3P f'(P)+f(P)
\be
g= \frac{A}{\rho}+ B \,  \frac{\asinh \rho}{\rho}.
\ee
The fall-off requirement in (\ref{pde1Hp}) implies $B=0$. The coefficient $A$ can be obtained by demanding that for $\rho \to 0$ one recovers the  flat space  Green's function. This sets $A=-1/(4 \pi)$. The Green's function is finally obtained by expressing $\rho$ in terms of $-Y \cdot Y'$ resulting in
\be
 \G(y;y') = -\frac{1}{4 \pi} \frac{1}{\sqrt{(Y \cdot Y')^2-1}}
\ee
\subsubsection{Green's function for Eq. (\ref{pde2Hp})}
We now seek for $G(y;\qh)$ such that
\be
(D^2+1)G=0, \quad G(\rho,\xh;\qh) \sim \frac{\ln \rho}{\rho}\delta^{(2)}(\xh,\qh'). \label{green2}
\ee
We just state the solution and verify it satisfies the desired conditions:
\be
G(y;\qh) =- \frac{1}{4 \pi} \frac{1}{Y \cdot q} \label{G}
\ee
where 
\be
Y \cdot q = -\sqrt{1+\rho^2} +\rho \xh \cdot \qh
 \ee
is the Minkowski product between $Y^\mu$  (\ref{Ymu}) and $q^\mu$  (\ref{qmu}). First, one can verify that 
\be
D^2 f(Y \cdot q) = (Y \cdot q)^2 f''(Y \cdot q)+3 (Y \cdot q)f'(Y \cdot q) \label{D2fHp}
\ee
from which it follows that $(Y \cdot q)^{-1}$ satisfies the desired equation. Next we note that
\be
G(y;\qh)  \sim \left\{ \begin{array}{lll} 1/\rho & {\rm for} & \xh \neq \qh \\ \rho & {\rm for} & \xh = \qh \end{array}\right.
\ee
and
\be
\int d^2\qh \, G(y;\qh) =  \frac{\asinh \rho}{\rho} \sim \frac{\ln \rho}{\rho}
\ee
from which it follows that $G$ satisfies the asymptotic behavior given in (\ref{green2}).\footnote{The reasoning here is the same as the one given in \cite{mcgreen}. The present case extends the analysis of \cite{mcgreen} from $n>1$ to $n=1$.}
\subsubsection{Relation between Green's functions and Eq. (\ref{phiphard})} \label{phiphardapp}
From the expressions above one can verify that the large $\rho$ behavior of $\G$ is dictated by $G$ according to\footnote{This is just a particular instance of more general relations between `bulk-bulk' and `bulk-boundary' Green's functions that are well known in the AdS/CFT literature.}
\be
\G(y;y') = -\frac{1}{\rho}G(y';\xh) +O(\rho^{-3}) \label{Gbb}
\ee
We now have all elements to show Eq. (\ref{phiphard}):
\ba
\phi(u=+\infty,\qh) &=& \lim_{\rho \to \infty} \rho \phone(\rho,\qh) \\
 &=& \lim_{\rho \to \infty} \rho \int d^3V' \G(\rho,\qh; y') j(y') \\
 &=&- \int d^3V' G(y';\qh) j(y') \\
 &=& \frac{1}{4\pi} \Qh_{\qh}.
\ea
The first equality arises from consistency of the null and time-infinity expansions. The second equality uses (\ref{solpoiss}) and the third (\ref{Gbb}).
In the last equality we used the expressions (\ref{Qhq}) and (\ref{G})
\subsection{Spatial infinity $\H^o$}\label{Hoapp}
Hyperbolic coordinates adapted to spatial infinity are
\be
\rho:= \sqrt{r^2 -t^2}, \quad \tau := \frac{t}{\sqrt{r^2-t^2}}
\ee
in terms of which the Minkowski line element reads:
\be
ds^2= d \rho^2 + \rho^2 d \sigma^2
\ee
with 
\be
d \sigma^2 = - \frac{d \tau^2}{1+\tau^2}+(1+\tau^2) q_{AB} dx^A dx^B = : h_{\alpha \beta} d y^\alpha d y^\beta
\ee
the line element of the unit hyperboloid $\Ho$.

For a field of the form
\be
\varphi(x) = \rho^{-n}\phn(y)
\ee
one has 
\be
\square \varphi(x) = \rho^{-n-2}(D^2 \phn-n(2-n) \phn), \label{boxrn}
\ee
where $D^2 \equiv h^{\alpha \beta}D_\alpha D_\beta$ is the wave operator on $\Ho$. By studying the differential operator on $\Ho$ defined by (\ref{boxrn})  one concludes:
\be
(D^2 -n(2-n) )\phn =0 \; \implies \; \phn  \sim 1/\t^n \quad {\rm or} \quad  \phn \sim 1/\t^{2-n}  \label{D2Ho}
\ee
when $\t \to \pm \infty$. Again, for $n=1$ these two asymptotic  solutions coincide but a new one appears:
\be
(D^2 -1) \phone =0 \; \implies \; \phone \sim 1/\t \quad {\rm or} \quad  \phone \sim \ln \t/\t  . \label{freeHo}
\ee
For the purposes of the present paper, we need the general solution associated to the each type of fall-offs:
\ba
(D^2 -1)\phone  &=  0  & {\rm with}  \quad \phone  \stackrel{\t \to + \infty}{\sim} \frac{1}{\t}  \phi_+(\xh) \label{pde1Ho} \\
(D^2 -1)\lamone & = 0 &{\rm with} \quad \lamone \stackrel{\t \to + \infty}{\sim} \frac{\ln \t}{\t}\lambda_+(\xh) . \label{pde2Ho}
\ea
For definitiveness we consider the `backwards' evolution problem of solving the wave equation with data on the asymptotic future. Note that, unlike the elliptic problem in $\Hp$, the source free equation (\ref{freeHo}) has two independent solutions. This corresponds to the fact that we now have a Cauchy problem and two data are needed to solve the equation (the field and its momentum). Asymptotically the two data correspond to the different fall-offs. 

We now describe the Green's function associated to  each equation. We will just state them and verify they satisfy the desired conditions. 
\subsubsection{Solution to Eq. (\ref{pde1Ho})} \label{pde1Hoapp}
The solution to  (\ref{pde1Ho}) we claim is given by:
\be
\phone(y) = \frac{1}{2\pi}\int d^2 \qh \, \delta(Y \cdot q) \phi_+(\qh). \label{phoneint}
\ee
First, we note that the analogue of Eq. (\ref{D2Hp}) in $\Ho$ is:
\be
D^2 f(Y \cdot q) = -(Y \cdot q)^2 f''(Y \cdot q)-3 (Y \cdot q)f'(Y \cdot q) . \label{D2fHo}
\ee
Noting the distributional identities $\sigma \delta'(\sigma) = -\delta(\sigma)$ and $\sigma^2 \delta''(\sigma)=2\delta(\sigma)$  one concludes that  $D^2 \delta (Y \cdot q)= \delta (Y \cdot q)$ and so (\ref{phoneint}) satisfies the desired differential equation. To check for the boundary condition,  note that the support of the Delta function in (\ref{phoneint}) is given by 
\be
Y \cdot q=0 \iff \xh \cdot \qh =\frac{\t}{\sqrt{1+\t^2}} \stackrel{\t\to +\infty}{=} 1-\frac{1}{2}\frac{1}{\t^2}+O(\t^{-4}).
\ee
If we denote by $\theta$ the angle between $\xh$ and $\qh$ then $\xh \cdot \qh \approx 1- \frac{1}{2}\theta^2$ for small $\theta$. Thus for large $\t$ the support of the Delta is a circle of radius $\t^{-1}$ centered around $\qh=\xh$. It then follows that  $\phone(y)$ defined by (\ref{phoneint}) satisfies $\phone(\t) \sim  \t^{-1}\phi_+(\xh)$ for  $\t \to +\infty$ as desired. The solution (\ref{phoneint}) also allow us to evaluate  $\phone$ in the $\t \to -\infty$ limit. In this case $\frac{\t}{\sqrt{1+\t^2}}\sim -(1-\frac{1}{2}\frac{1}{\t^2}) $ and the Delta becomes supported in a $|\t|^{-1}$ radius circle centered at $\qh=-\xh$. This then implies relation (\ref{antipphi}).

\subsubsection{Solution to Eq. (\ref{pde2Ho})}
The solution to  (\ref{pde2Ho}) we claim is given by:
\be
\lamone(y) =  -\frac{1}{4\pi}\int d^2 \qh \, \frac{\lambda_+(\qh)}{Y \cdot q} , \label{lamoneint}
\ee
which is the $\Ho$ version of the analogous expression found in the $\Hp$ case. The proof then goes parallel to that case. The  differential equation can be seen to be satisfied by using Eq. (\ref{D2fHo}). For the boundary condition we note that
\be
\frac{1}{Y \cdot q}  \stackrel{\t \to \pm \infty}{=} \left\{ \begin{array}{lll} O(1/\t) & {\rm for} & \xh \neq \pm \qh \\ O(\t) & {\rm for} & \xh =\pm \qh \end{array}\right.
\ee
and
\be
-\frac{1}{4\pi}\int  \frac{d^2\qh}{Y \cdot q} =  \frac{\asinh \t}{\sqrt{1+\t^2}} \stackrel{\t \to \pm \infty}{\sim} \pm \frac{\ln |\t|}{|\t|} .
\ee
from which the $\t \to + \infty$ condition in (\ref{pde2Ho}) follows. The $\t \to - \infty$ limit yields the `antipodal matching'  $\lambda_-(\xh)=-\lambda_+(-\xh)$.

\subsubsection{Relation between fields (\ref{pde1Ho}) and (\ref{pde2Ho}) } \label{relHoapp}
The solutions $\phone(y)$ and $\lamone(y)$ given in (\ref{phoneint}) and (\ref{lamoneint}) represent two type of solutions of the  differential equation
\be
(D^2 -1)f = 0. \label{kgf}
\ee
A geometrical characterization of each type of solution can be given by considering the inversion map on $\Ho$:
\be
y^\alpha \to -y^\alpha :=(-\t, -\sqrt{1+\t^2} \xh) \quad \quad  (Y^\mu \to - Y^\mu).
\ee
From the integral representations  (\ref{phoneint}) and (\ref{lamoneint}) it  follows that each type of field is of definite parity:
\be
\phone(-y) = \phone(y) \quad {\rm and} \quad \lamone(-y)=-\lamone(y).
\ee
%That is, the `$\phi$-type' fields are even and the `$\lambda$-type' fields are odd. 
A general solution to (\ref{kgf}) will in general be a sum of the two type of solutions. Thus if we denote by  $\Gamma$ the space of all fields satisfying (\ref{kgf}) we have:
\be
\Gamma =\Gamma^{\rm even} \oplus \Gamma^{\rm odd}. \label{Gamsplit}
\ee
We now show that in fact  (\ref{Gamsplit}) corresponds to a ``$q-p$'' splitting when  $\Gamma$ is regarded as a phase space. Recall that fields  satisfying a (massive) Klein-Gordon equation such as (\ref{kgf}) %can be thought of as arising from an action  of (unit mass) Klein-Gordon fields on $\Ho$. As such, it 
form a symplectic vector space  under the usual Klein-Gordon symplectic product
\be
\O(f,g)= \int_{C} dS_\alpha \sqrt{h}h^{\alpha \beta}(\partial_\beta f g-\partial_\beta g f), \label{Okg}
\ee
where $C$ is any Cauchy surface on $\Ho$.  We claim that
\be
\O(\phone,\phone')= \O(\lamone,\lamone')=0, \label{Ozero}
\ee
\be
\O(\phone,\lamone)=\int_{S^2} d^2 V \lambda_+(\xh) \phi_+(\xh).
\ee
These results can be obtained by evaluating (\ref{Okg}) in the $\t \to \infty$ slice and using the asymptotic forms of the fields. It is  interesting however to see how Eq. (\ref{Ozero}) arises for the  $\t=0$ slice (recall $\Omega$ is independent of the choice of Cauchy slice). 

 From the $\t=0$ slice perspective,  $\Gamma$ is the space of initial conditions $(f,\dot{f})$  on the $\t=0$ sphere with symplectic product 
 \be
 \Omega((f,\dot{f}),(g,\dot{g})) = \int_{S^2} d^2V (\dot{g} f - \dot{f} g). \label{Otz}
 \ee
By evaluating the $\phi$ and $\lambda$ integral solutions (and their $\t$ derivatives) at $\t=0$ one concludes that
\be
(f, \dot{f}) \in \Gamma^{\rm even} \iff  f \text{ even  }, \dot f  \text{ odd}
\ee
\be
(f, \dot{f}) \in \Gamma^{\rm odd} \iff  f \text{ odd  }, \dot{f} \text{ even}
\ee
where the parity of $f,\dot f$ refers  to $S^2$. Since an odd function on the sphere integrates to zero, it follows that (\ref{Otz}) vanishes if evaluated on either $\Gamma^{\rm even}$ or $\Gamma^{\rm odd}$.

We finally note that the $\t=0$ sphere corresponds to the limiting spheres of $t=$constant spacetime slices $\Sigma_t$ ($t=$ Minkowski time). From this perspective, $\Gamma^{\rm even}$ determines the asymptotic parity  of $\varphi$-initial data  on $\Sigma_t$. These  are the scalar field analogue of the Regge-Teitelboim parity conditions in gravity \cite{rt}.

\section{Asymptotics of $\Lambda(x)$ at null infinity} \label{lamapp}
Here we describe the null infinity behavior of $\Lambda$ as given in Eq. (\ref{lambulk}). Rather than attempting a direct evaluation of (\ref{lambulk}) in the $r \to \infty, u=$const. limit, we will make use of the results from the previous appendix and work in hyperbolic coordinates as an intermediate step.  Consider first hyperbolic coordinates adapted to future time infinity.  (these only  allow us to cover the $u>0$ region of $\I^+$).  In these coordinates we have:
\be
\Lambda(x) = \frac{\lamhp(y)}{\t} = \frac{1}{\rho \t}( \ln \rho \lambda(\xh) +O(\rho^{-1})).
\ee
We now write this expression in retarded $u,r$ coordinates. Noting that $\rho \t= r$ and
\be
\rho = \frac{r}{\sqrt{2 u r+u^2}}= \sqrt{\frac{r}{2 u}}  \left(1+O(r^{-1}) \right),
\ee
\be
\ln \rho= \frac{1}{2}\ln r-\frac{1}{2} \ln (2 u)+O(r^{-1})
\ee
one finds
\be
\Lambda(x)= \frac{1}{2 r}\lambda(\xh) (\ln r - \ln (2 u) +O(r^{-1/2}))  \quad \quad (u>0). \label{upos}
\ee
We recall the above expression is only valid for $u>0$. To cover the $u<0$ region we consider the expression of $\Lambda(x)$ in hyperbolic coordinates adapted to spatial infinity:
\be
\Lambda(x) = \frac{\lamho(y)}{\rho} = \frac{1}{\rho \t}( \ln \t \lambda(\xh) +O(\t^{-1})).
\ee
In terms of $(r,u)$ coordinates we now have $\rho \t = r+ u$ and
\be
\t = \frac{r+u}{\sqrt{-2 u r-u^2}}= \sqrt{\frac{r}{-2 u}}  \left(1+O(r^{-1}) \right),
\ee
leading to
\be
\Lambda(x)= \frac{1}{2 r}\lambda(\xh) (\ln r - \ln (-2 u) +O(r^{-1/2}))  \quad \quad  (u<0). \label{uneg}
\ee
Combining (\ref{upos}) and (\ref{uneg}) we obtain (\ref{lamnull}).

\subsection{Eq. (\ref{qnullfin})}
We concluded that near null infinity $\Lambda(x)$ has an expansion of the form
\be
\Lambda(x) = \frac{\ln r}{r} \, \lamln(u,\xh) + \frac{1}{r} \, \lamo(u,\xh) + O(r^{-1-\epsilon}) \label{lamnulapp}
\ee
with 
\be
 \lamln(u,\xh) = \frac{1}{2} \lambda(\xh), \quad \lamo(u,\xh)= -\frac{1}{2} \ln |2u|  \lambda(\xh). \label{lamsnull}
\ee
From (\ref{lamnulapp}) one finds the following null infinity expansion for $k^{ru}$ (\ref{kab}): 
\be
k^{ru}= -  (\ln r \lamln + \lamo) u \partial_u \phisp + (u \partial_u \lamo - \lamln) \phisp + O(r^{-\epsilon}) \label{krunull}
\ee
The $u \to \pm \infty$ fall-offs for $\phisp$ are such that $\partial_u \phisp=O(u^{-1-\epsilon})$. This implies   the first term in (\ref{krunull})  vanishes when  $u \to \pm \infty$ and so it does not contribute to the charge (\ref{qnullfirst}) (in particular the potential logarithmic divergence is absent). From (\ref{lamsnull}) one  finds the second term is
\be
(u \partial_u \lamo - \lamln)  \phisp= -  \phisp(u,\xh) \lambda(\xh),
\ee 
from which  (\ref{qnullfin}) follows.

\section{Shift symmetry for free massless fields} \label{ss3}
In this appendix we discuss  `shift symmetries' of the free massless scalar theory. We will discuss in parallel the free Maxwell and linearized gravity cases in harmonic gauge to highlight their similarities. 

As described below, the   naturally occurring `shift' field for the massless spin $s=0,1,2$ theory   can be written in the  simple form,
\be
\Lambda^{(s)}(x) := -\frac{(x \cdot x)^s}{4 \pi}\int d^2 \qh \, \frac{\lambda(\qh)}{\, (q \cdot x)^{s+1}} \, . \label{Lams}
\ee
 $\Lambda^{(s)}(x)$ depends on a function on the sphere $\lambda(\qh)$ which  determines its  asymptotic value.\footnote{One can also consider  Poincare transformed  versions of (\ref{Lams}). These are however redundant from the perspective of infinity, namely  $\Lambda_\lambda(L^\mu_\nu x^\nu + b^\mu) \to\Lambda_{\lambda^L}(x)$ with $\lambda^L(\qh)$ a Lorentz transformed  version of $\lambda(\qh)$.}  It satisfies  the scaling behavior,
\be
t^{(1-s)}\Lambda^{(s)}(t x) = \Lambda^{(s)}(x), \label{scaleinvlam}
\ee
 and obeys the free wave equation
\be
\square \Lambda^{(s)}=0.
\ee

%that is, it is scale invariant with  scaling dimension $(1-s)$.

For $s=0$ (\ref{Lams}) becomes the expression for the field $\Lambda(x)$ introduced in Eq. (\ref{lambulk}).  For $s=1$ (\ref{Lams}) corresponds to the large $U(1)$ gauge parameter in harmonic gauge given in   \cite{clmass}  which satisfies
\be
\Lamo \to \lambda(\xh)
\ee
at null infinity.
Finally, for $s=2$ the vector field defined by $\xi_a := \partial_a \Lamt$ corresponds to the supertranslation vector field in harmonic gauge given  in \cite{clmass2}  which satisfies
\be
\xi^a \to \lambda(\xh)  \partial_u
\ee
at null infinity. These three parameters yield the  following family of `shifted vacua'  for the  scalar, Maxwell and gravity theories: 
\ba
\varphi & = & \Lamz  \label{vacph} \\
A_\mu & = & \partial_\mu \Lamo  \label{vacA} \\
h_{\mu \nu} & = & 2 \partial_\mu \partial_\nu \Lamt. \label{vach}
\ea
Condition  (\ref{scaleinvlam}) implies all three solutions are scale  invariant (with the standard scaling dimension 1). %The significance of this property is not yet fully understood to us.

So far we have tried to exhibit the similarities between the different massless theories. We now discuss the differences between the $s=0$ and $s=1,2$ cases. First, the configurations (\ref{vacA}), (\ref{vach}) correspond to  nontrivial vacuum configurations associated to the large local symmetry group.\footnote{To simplify the discussion we are ignoring ``superrotated''  vacua in the gravity case.} In particular, all these solutions have zero energy. One may be tempted to interpret (\ref{vacph}) as a non-trivial vacuum solution. However if one tries to compute the energy associated to this configuration one gets infinity!\footnote{This can be achieved, for instance, by evaluating the total energy near  null infinity using the expansion (\ref{lamnull}).} Related to this is the fact that the transformation generated by the `soft theorem charge' does not yield a solution of the type (\ref{vacph})   but rather one of the form given in Eq. (\ref{dellamphi}) (which does have zero energy). 

As discussed earlier, in the gauge and gravity cases the currents associated to the `shift' transformations leading to (\ref{vacA}) and (\ref{vach}) are total derivatives as in (\ref{jk}). This is not so in the scalar case. The shift transformation
\be
\delta_\Lambda \varphi = \Lambda 
\ee
with $\square \Lambda=0$ is a symmetry of the  free $\varphi$ Lagrangian with  Noether current given by:
\be
j^a[\Lambda] = \sqrt{\eta}( \varphi \nabla^a \Lambda - \Lambda \nabla^a \varphi) \label{jS}
\ee
 which is not of the total derivative form.  One may ask if there is any relation between (\ref{jS}) and the total derivative current (\ref{kab}) . If we denote by
\be
\delta_X \varphi = \varphi + X^a \partial_a \varphi
\ee
the action of dilatations on $\varphi$ then one finds  (in the free theory case)
\be
\delta_X j^a[\Lambda]  = - \partial_b k^{ab},
\ee
with  $k^{ab}$ given in Eq. (\ref{kab}). %Further elucidation of these structures is left for the future. 

\end{document}